\documentclass[aps, prl, reprint, superscriptaddress, nofootinbib]{revtex4-1}
\usepackage[utf8]{inputenc}
\usepackage[english]{babel}
\usepackage{physics}
\usepackage{times}
\usepackage{graphicx,epsfig}
\usepackage{color}
\usepackage[usenames,dvipsnames]{xcolor}
\definecolor{linkcolor}{rgb}{0,0,0.6}		
\definecolor{bleu}{HTML}{1732a6}
\usepackage[colorlinks=true, pdfstartview=FitV, linkcolor= linkcolor, citecolor= linkcolor, urlcolor= linkcolor, hyperindex=true, hyperfigures=false]{hyperref}

\addto\captionsenglish{}

\newcommand*{\e}{\text{e}}
\renewcommand*{\d}{\text{d}}
\newcommand*{\ii}{\text{i}}
\newcommand*{\bin}{\ensuremath{\hat{b}_\text{in}}}
\newcommand*{\bout}{\ensuremath{\hat{b}_\text{out}}}
\newcommand*{\bbin}{\ensuremath{{b}_\text{in}}}

\newcommand*{\E}{\mathcal{E}}

\begin{document}

\title{The energetic cost of work extraction}
\author{Juliette Monsel}
\affiliation{Univ. Grenoble Alpes, CNRS, Grenoble INP, Institut N\'eel, 38000 Grenoble, France}
\author{Marco Fellous-Asiani}
\affiliation{Univ. Grenoble Alpes, CNRS, Grenoble INP, Institut N\'eel, 38000 Grenoble, France}
\author{Benjamin Huard}
\affiliation{Univ Lyon, ENS de Lyon, Univ Claude Bernard, CNRS, Laboratoire de Physique, Lyon, France}
\author{Alexia Auff\`eves}
\email{alexia.auffeves@neel.cnrs.fr}
\affiliation{Univ. Grenoble Alpes, CNRS, Grenoble INP, Institut N\'eel, 38000 Grenoble, France}
\begin{abstract}
We analyze work extraction from a qubit into a wave guide (WG) acting as a battery, where work is the coherent component of the energy radiated by the qubit. The process is stimulated by a wave packet whose mean photon number (the battery's charge) can be adjusted. We show that the extracted work is bounded by the qubit's ergotropy, and that the bound is saturated for a large enough battery's charge. If this charge is small, work can still be extracted. Its amount is controlled by the quantum coherence initially injected in the qubit's state, that appears as a key parameter when energetic resources are limited. This new and autonomous scenario for the study of quantum batteries can be implemented with state-of-the-art artificial qubits coupled to WGs.
\end{abstract}

\maketitle

A central part of thermodynamics consists in designing protocols to extract energy from physical systems, without extracting their entropy. These key sequences are called ``work extraction", the physical systems being ``working substances". Work can then be immediately used, or stored into a battery. In the quantum realm, such protocols have been fruitfully modeled by unitary operations, the maximal amount of extractable work defining the so-called ergotropy of the working substance \cite{Allahverdyan_2004}. In quantum optics, the maser provides a paradigmatic example of such sequence \cite{Scovil59}. Namely, a qubit (the working substance) provides work to a resonant electromagnetic mode (the battery) by stimulated emission. The battery is initially charged with a coherent field containing a large number of photons, a large enough initial charge ensuring that the entanglement between the field and the qubit remains negligible under free evolution. Therefore the reduced qubit's evolution can be safely taken as unitary. Because of its conceptual simplicity, work extraction by stimulated emission remains the core mechanism to analyze the performances of quantum heat engines, both theoretically \cite{ScullyScience03, Uzdin15} and experimentally \cite{Klatzov2019}. It is also at play in recent implementation of quantum Maxwell's demons \cite{Masuyama18,Naghiloo18}, where work is inferred from energy measurements performed on the working substance. Direct and experimentally feasible strategies to evidence work extraction, based on the measurement of the battery itself, are thus highly desirable. 

This is a strong motivation for investigating resonant work extraction in quantum batteries, an emerging topic that currently attracts great interest \cite{QuThermo}. Work extraction was shown to be affected by the charging dynamics \cite{hovhannisyan_entanglement_2013, Quantacell}, the quantum correlations between the working substance and the battery \cite{andolina_charger-mediated_2018, Andolina2019}, the collective effects between the subsystems forming the battery \cite{alicki_entanglement_2013, hovhannisyan_entanglement_2013, Quantacell, Kavan2017, Campisi2018, le_spin-chain_2018}. It was first studied in an abstract way, the battery being a collection of identical quantum systems charged by an external time-dependent operator \cite{alicki_entanglement_2013, hovhannisyan_entanglement_2013, Quantacell, Kavan2017}. Then, more concrete systems, such as qubits in a cavity \cite{Campisi2018, Andolina2019} or spin-chains \cite{le_spin-chain_2018}, were considered. Most studies focus on the maximization of either the charging power \cite{alicki_entanglement_2013, hovhannisyan_entanglement_2013, Quantacell, Kavan2017, andolina_charger-mediated_2018, Campisi2018, le_spin-chain_2018} or the ergotropy \cite{alicki_entanglement_2013, hovhannisyan_entanglement_2013, Quantacell, Andolina2019}.

In this Letter, we take another standpoint and study how the initial quantum coherence present in the working substance, as well as the initial charge of the battery impact work extraction. The working substance is a qubit embedded into a wave guide (WG), i.e. a reservoir of electromagnetic modes that acts as a battery. Work (resp. heat) is defined as the coherent (resp. incoherent) fraction of energy radiated by the qubit in the battery. The emission process is stimulated by a resonant wave packet propagating in the WG, the mean photon number it contains defining the initial (and adjustable) charge of the battery.  Studying this scenario offers a number of advantages. Firstly, it provides a new and autonomous scenario for the study of quantum batteries coupled to working substances by energy conserving transformations. Secondly, it matches the textbook situation of work extraction by stimulated emission, in the limit of large number of photons. Finally, it corresponds to a realistic experimental framework dubbed WaveGuide Quantum ElectroDynamics (WG-QED) \cite{Kimble,Daniel12} that is routinely implemented both in superconducting  \cite{Gu2017} and semiconducting circuits \cite{Few-photons,Loredo2019,Lodahl19}.\\

We first show that the qubit's ergotropy is an upper bound for work extraction, and that this bound is saturated in the limit where the reduced qubit's evolution is unitary. The price to reach this bound is thus a large initial battery's charge, which corresponds to heavy energetic resources. We then consider the case of an initially uncharged battery. Originally, work can be spontaneously extracted even though the mechanism is dissipative. The amount of extracted work scales like the quantum coherence initially injected in the qubit's state. We finally consider the case of a battery of intermediate charge. The battery's energy and the qubit's coherence appear as complementary resources to optimize work extraction. These results reveal that quantum coherence is bound to play a key role to control energetic transfers with limited energetic supplies.\\

\begin{figure}[htb]
  \includegraphics[width=\linewidth]{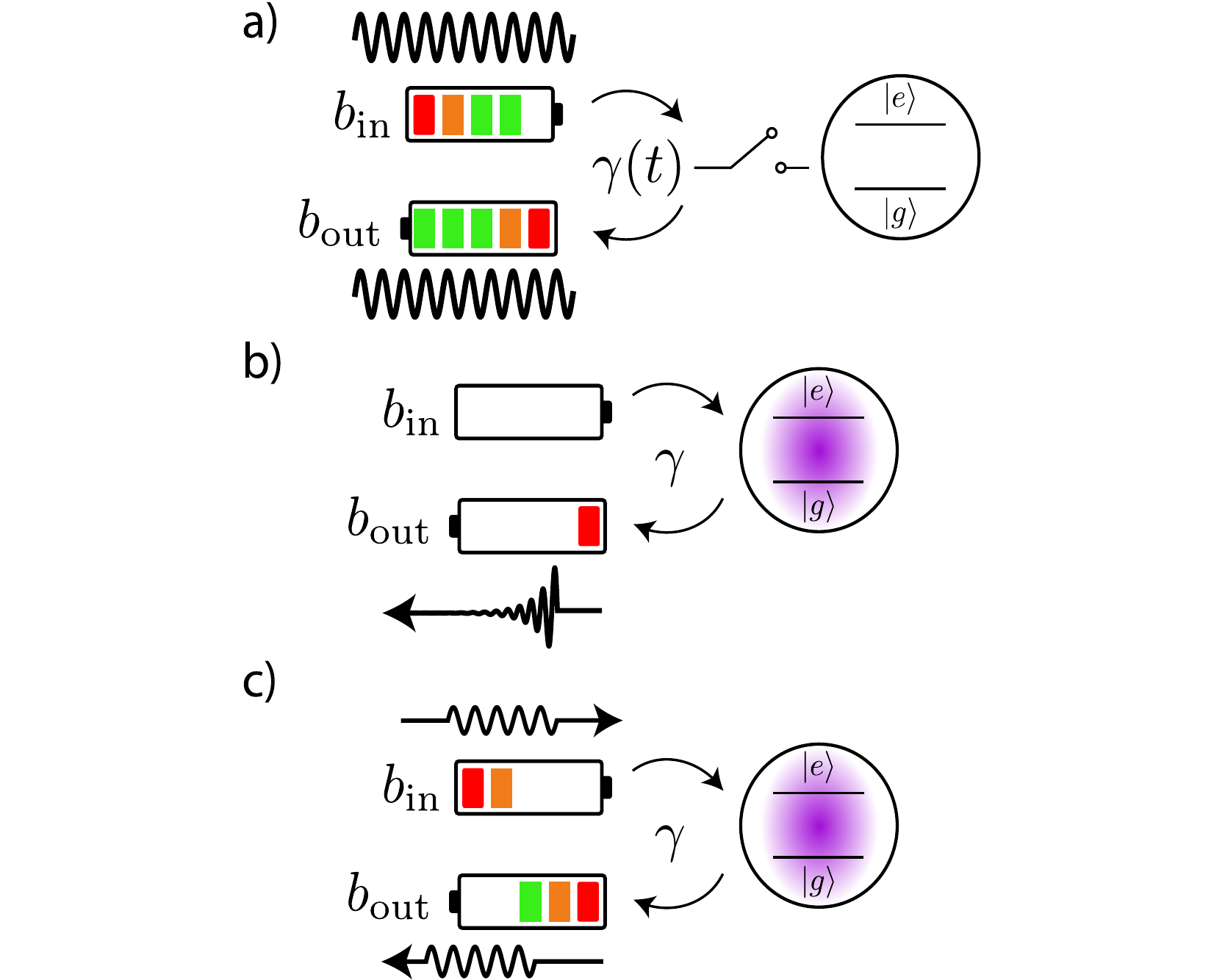}
  \caption{\label{fig1} Three different scenarios of work extraction from a qubit (working substance) into a WG (battery). The spontaneous emission rate is denoted $\gamma(t)$, the input photon rate $\dot{N}(t) = |b_\text{in}(t)|^2$. The extracted work rate reads $\dot{W} =\hbar \omega_0 (|b_\text{out}(t)|^2-|b_\text{in}(t)|^2)$, see text. \textbf{(a)} The input photon rate is constant, the qubit-WG coupling is switched off after $\tau_\text{opt}$:  $\dot{N}(t) = \dot{N}$, $\gamma(t)=\gamma$ for $t \in [0,\tau_\text{opt}[$, $\gamma(t)=0$ for $t\geq \tau_\text{opt}$. \textbf{(b)} The WG is initially uncharged, the qubit-WG coupling is constant: $\dot{N}=0$, $\gamma(t)=\gamma$ for all $t$. \textbf{(c)} The qubit-WG is constant, the WG charged with a wave packet of duration $\tau$: $\gamma(t)=\gamma$, $\dot{N}(t) = \dot{N}$ for $t \in [0,\tau[$, $\dot{N}=0$ for $t\geq \tau$. }
\end{figure} 

The setup under study involves a qubit (working substance) and a WG (battery) as depicted in Fig.~\ref{fig1}. The qubit's excited (resp. ground) state is denoted $\ket{e}$ (resp. $\ket{g}$), its transition frequency $\omega_0$. The WG is a reservoir of electromagnetic modes, an initially empty battery corresponding to all modes set at zero temperature. In this case, the qubit-WG coupling induces the spontaneous emission rate $\gamma$. This coupling can possibly be switched off by an external operator, such that $\gamma=0$. Conversely, the battery can initially be filled with a coherent input field of complex amplitude $b_\text{in}(t)$ resonant with the qubit's frequency. In the input-output formalism, $|b_\text{in}(t)|^2=\dot{N}(t)$ (resp. $P_\text{in}=\hbar \omega_0 \dot{N}$) corresponds to the input photon rate (resp. the input power), allowing to define the initial battery's charge as $\bar{N}= \int_0^\infty \dd  t  |b_\text{in}(t)|^2$.  The Hamiltonian ruling the qubit's evolution is  $\hat{H}(t) = \hat{H}_0 + \hat{H}_\d (t)$ where $\hat{H}_0 = \frac{\hbar \omega_0}{2}(\hat{\sigma}_z+1)$ is the free Hamiltonian and $\hat{H}_\d(t) =\ii\hbar \sqrt{\gamma(t) \dot{N}(t)} (\hat{\sigma}_- \e^{\ii\omega_0 t} - \hat{\sigma}_+ \e^{-\ii\omega_0 t})$ the drive Hamiltonian. We have introduced $\hat{\sigma}_z = \dyad{e}-\dyad{g}$, $\hat{\sigma}_- = \ket{g}\bra{e} = \hat{\sigma}^\dagger_+$ and $\gamma(t) \in \{ \gamma, 0\} $, depending whether the WG is coupled or not to the qubit. Denoting the dissipator as ${\cal D}[O]\rho = O \rho O^\dagger - \{O^\dagger O, \rho \}/2$, the evolution of the qubit's state $\rho(t)$ obeys the Lindblad equation
\begin{equation} \label{bloch}
\dot{\rho} = -\frac{\ii}{\hbar} [\hat{H}(t),\rho] + \gamma(t) {\cal D}[\hat{\sigma}_-]\rho.
\end{equation}
Conversely, the output field operator $\hat{b}_\text{out}$ and the output power $P_\text{out} = \hbar \omega_0 \langle \hat{b}^\dagger_\text{out} \hat{b}_\text{out} \rangle (t)$ verify (See Suppl.~\cite{Suppl})
\begin{align} 
\hat{b}_\text{out}(t) &= b_\text{in}(t) + \sqrt{\gamma} \hat{\sigma}_-(t),  \label{bout} \\
P_\text{out}(t) &= P_\text{in}(t) - \dot{{\cal E}}(t). \label{Pout}
\end{align}
${\cal E}(t) = \Tr[\rho(t) \hat{H}(t)]$ stands for the qubit's mean energy. As expected, the battery collects all the power radiated by the qubit (See Eq.~\eqref{Pout}). Owing to the WG geometry enabling the efficient detection of light, the power change resulting from the qubit-field interaction can be measured in state-of-the-art superconducting and semi-conducting devices \cite{Delsing12,Walraff12,Few-photons,Lodahl19}. More originally with respect to recent studies \cite{QuThermo, Quantacell, Kavan2017, Campisi2018, Barra, Andolina2019}, our battery also plays the role of a drive, that induces a time-dependent Hamiltonian on the working substance and performs work on it. The present framework captures the three scenarios for work extraction illustrated in Fig.~\ref{fig1} that will be considered.
In the case (i) (Fig.~\ref{fig1}a), the WG's state is characterized by some constant photonic rate $\dot{N}$. An external operator controls the duration of the qubit-WG interaction, $\gamma(t)=\gamma$ for $t \in [0,\tau_\text{opt}[$, $\gamma(t)=0$ for $t\geq \tau_\text{opt}$.  In the case (ii) (Fig.~\ref{fig1}b), the qubit is constantly coupled to the uncharged WG: $\gamma(t) = \gamma$ and $\dot{N}=0$. In the case (iii) (Fig.~\ref{fig1}c), the qubit is constantly coupled to the WG charged with a wave packet of duration $\tau$: $\gamma(t)=\gamma$ and $\dot{N}(t) = \dot{N}$ for $t \in [0,\tau[$, $\dot{N}=0$ for $t\geq \tau$. In contrast to (i), the cases (ii) and (iii) are ``autonomous" in the sense that no external operator is involved.\\

\begin{figure}[htb]
    \includegraphics[width=\linewidth]{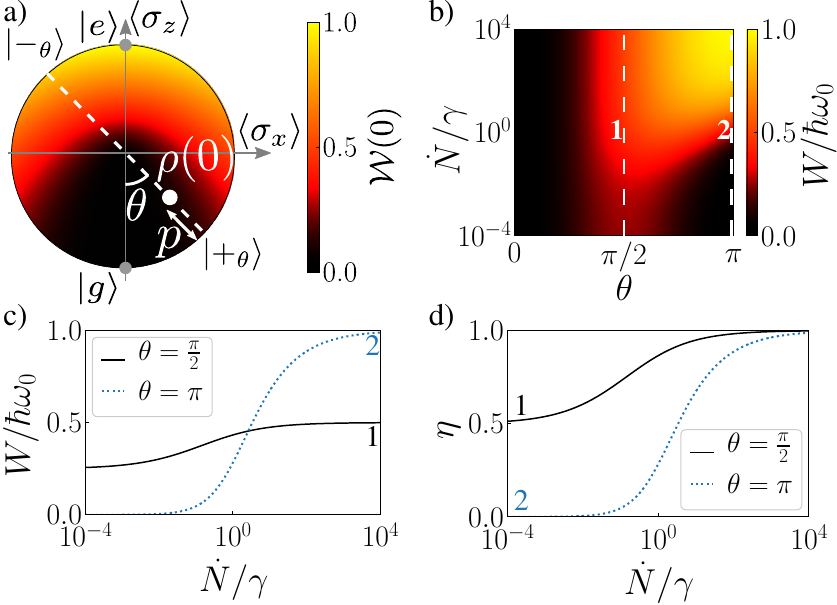}
    \caption{\label{fig2}
 Continuous regime of work extraction (Case (i), see text). \textbf{(a)} Ergotropy ${\cal W}(0)$ of the qubit's initial state $\rho(0) = (1-p) \dyad{+_\theta} + p\dyad{-_\theta}$ in the Bloch representation.   \textbf{(b)}  Maximal work extracted $W_\text{opt}$ after the optimal coupling time $\tau_\text{opt}$, as a function of the input photon rate $\dot{N}$ (in units of $\gamma$) and $\theta$, for a qubit's initial state $\rho(0) = \dyad{+_\theta}$.  \textbf{(c)}  $W_\text{opt}$ as a function of $\dot{N} / \gamma$ for $\theta = \pi/2$ (solid black) and $\theta=\pi$ (dotted blue).  \textbf{(d)}  $\eta$ as a function of $\dot{N} / \gamma$ for $\theta = \pi/2$ (solid black) and $\theta=\pi$ (dotted blue). }
\end{figure}

In all scenarios, the qubit is initially prepared in the quantum state $\rho(0)= p \dyad{-_\theta} + (1-p) \dyad{+_\theta}$, where $p \in [0,1/2]$, $\theta \in [0,\pi]$ with $\ket{+_\theta} = \sin(\theta/2) \ket{e} +\cos(\theta/2) \ket{g}$ and $\ket{-_\theta} = -\cos(\theta/2) \ket{e} + \sin(\theta/2) \ket{g}$. This initial state can be prepared experimentally by means of unitary (Rabi oscillations) or non-unitary operations (bath engineering techniques \cite{harrington_bath_2018}). It is characterized by its mean energy ${\cal E}(0) = \text{Tr}[\rho(0)\hat{H}_0]$, its coherence in the energy basis $s(0)= \Tr[\rho(0) \hat{\sigma}]$, and its ergotropy ${\cal W}(0)$. As stated above, the ``ergotropy" of a quantum state is the maximal amount of energy that can be extracted by unitary operations \cite{Allahverdyan_2004}, a positive (resp. null) ergotropy defining active (resp. passive) states. The qubit's ergotropy equals in the present case ${\cal W}(0) = \hbar\omega_0 (1 - 2p)\sin[2](\theta/2)$ (See Fig.~\ref{fig2}a and \cite{Suppl}) and obviously verifies ${\cal W}(0) \leq {\cal E}(0)$. As passive states, the thermal states ($ \langle \hat{\sigma}_z \rangle \in [-1,0], \langle \hat{\sigma}_x \rangle = \langle \hat{\sigma}_y \rangle =0$) contain no ergotropy. Reciprocally pure states verify ${\cal E}(0)={\cal W}(0)$, meaning that all the qubit's energy can be extracted unitarily. At $t=0$, the qubit is coupled to the WG through Eq.~\eqref{bloch}. \\

We now precise how to assess the quality of the work extraction from the qubit into the WG. In recent proposals \cite{Andolina2019, Barra, QuThermo}, the battery's ergotropy has been chosen as the proper quantity to maximize. However by definition, extracting the ergotropy of a quantum system like the battery requires the ability to perform unitary operations. As we show below, such ability consumes heavy energetic resources. Here we rather choose to  optimize the preparation of some directly useful state of the battery.  Being more specific, the battery prepared in such state should be able to perform a thermodynamic work on another quantum system without any further transformation. As seen above, such state simply corresponds to a coherent state. We shall thus define as ``work" (resp. ``heat") the energy carried by the coherent (resp. incoherent) component of the field radiated by the qubit in the WG. Introducing $b_\text{out}(t) = \langle \hat{b}_\text{out}(t) \rangle$, the work rate reads $\dot{W} = \hbar \omega_0 (|b_\text{out}(t)|^2 - |b_\text{in}(t)|^2)$, yielding (See Suppl.~\cite{Suppl})
\begin{align} 
\dot{W}(t)  =& \hbar \omega_0(\gamma |s(t)|^2 + \Omega \Re(s(t)e^{\ii \omega_0 t})), \label{eqW} \\
\dot{Q}(t) =& \hbar \omega_0 \gamma (P_e(t) - |s(t)|^2) \label{eqQ}
\end{align}
$\dot{Q}(t)$ stands for the heat rate and by energy conservation $-\dot{{\cal E}}(t) = \dot{W}(t)  + \dot{Q}(t)$. $P_e(t)$ is the population of the excited level and the qubit's dipole reads $s(t) =\Tr[\rho(t) \hat{\sigma} ]$, such that $\dot{Q}\geq 0$. We now investigate how to optimize work extraction in the three scenarios listed above and in Fig.~\ref{fig1}.\\

We first consider the scenario (i). The qubit-WG coupling time $\tau_\text{opt}$ is externally adjusted to maximize the amount of extracted work, $W_\text{opt}= \int_0^{\tau_\text{opt}}\dot{W}\dd t$. $W_\text{opt}$ is plotted in Fig.~\ref{fig2}b as a function of $\dot{N}/\gamma$ and $\theta$, for a pure qubit's initial state $\rho(0) = \dyad{+_\theta}$. For fixed $\theta$, $W_\text{opt}$ increases with $\dot{N}$ and reaches a maximal value when $\dot{N} \gg \gamma$ (See Fig.~\ref{fig2}c). This condition corresponds to the stimulated regime of light-matter interaction, where the dissipation induced by spontaneous emission captured by the Lindbladian of Eq.~\eqref{bloch} is negligible. Therefore $\dot{Q}=0$ and all the energy radiated by the qubit corresponds to work funneled into the driving mode. In this limit, the qubit-WG interaction is a unitary minimizing the qubit's energy, yielding $W_\text{opt}={\cal W}(0)$ by definition of the ergotropy. A maximal work extraction is obtained for $\theta = \pi$ and $\tau_\text{opt}=\pi/ 2\sqrt{\gamma \dot{N}}$ realizing a $\pi$-pulse, which describes a single photon amplifier \cite{sotier_femtosecond_2009}. \\

More generally, the initial qubit's ergotropy is an upper bound for work extraction, allowing to define a yield for the protocol $\eta = W/{\cal W}(0)$ (See Fig.~\ref{fig2}d and \cite{Suppl} for a general demonstration). This is the first result of this paper. As shown above, the bound is saturated in the limit of unitary operations. Interestingly, our framework reveals that this limit requires heavy energetic supplies. Even if we only consider the number of photons having interacted with the qubit, $N_\text{int}=\dot{N}\tau_\text{opt}$, this limit corresponds to a very large $N_\text{int}$. As explained above, the qubit evolves unitarily when $\dot{N}\gg \gamma$, which occurs when $N_\text{int}\approx  \sqrt{\dot{N}/\gamma}\gg 1$. Conversely, it appears in Fig.~\ref{fig2}b and c that a non-negligible work extraction is possible for $\dot{N}\leq \gamma$. This can be realized if $\theta \sim \pi/2$, i.e. provided some coherence is initially injected in the qubit's state. Quantum coherence thus appears as a key parameter, that compensates for a weak battery's charge and can therefore be fruitfully used if energetic resources are limited. \\

\begin{figure}[htb]
    \includegraphics[width=\linewidth]{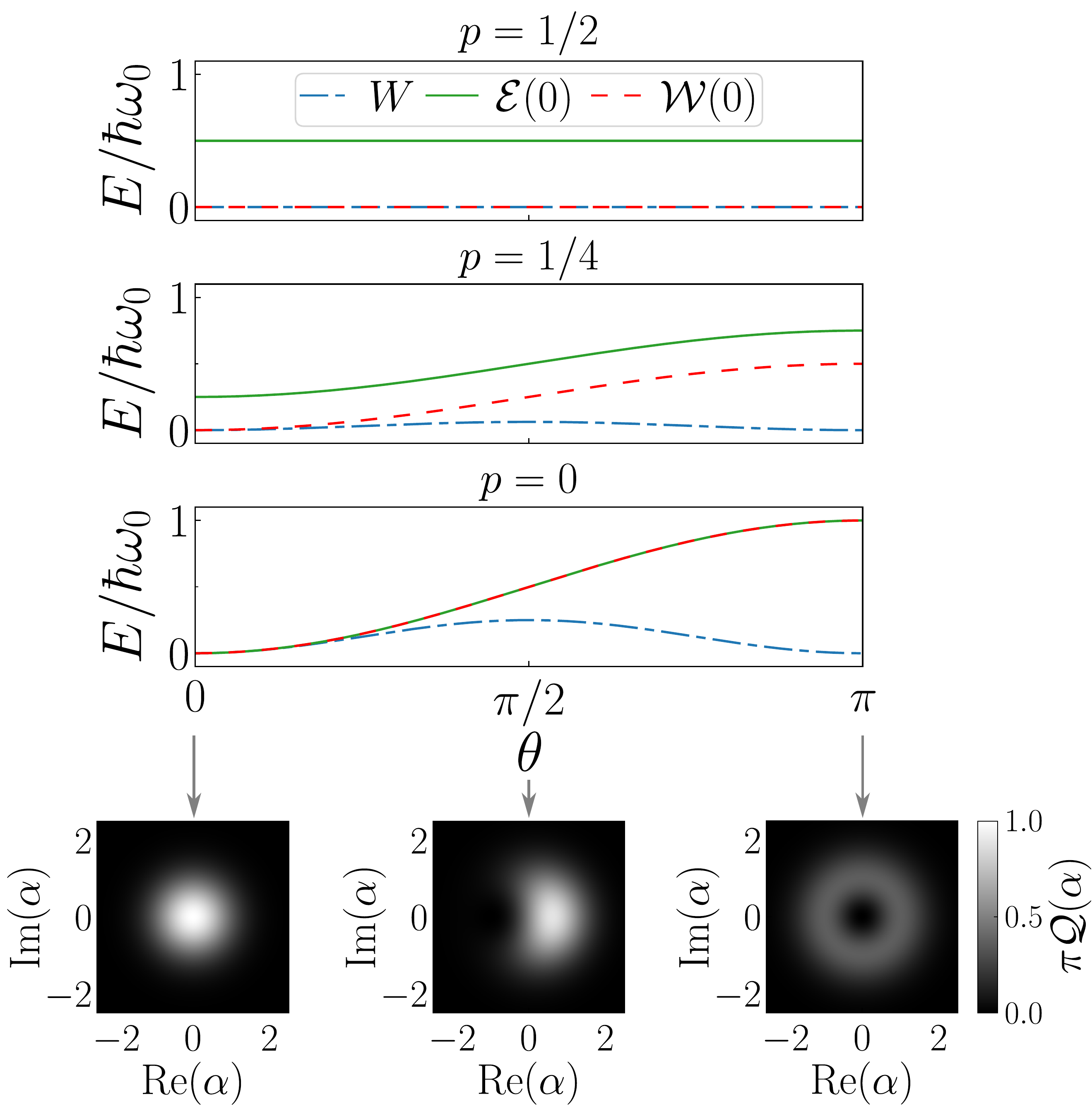}
    \caption{\label{fig3}
        Spontaneous regime of work extraction (Case (ii), see text). Initial energy ${\cal E}(0)$ (solid green), ergotropy ${\cal W}(0)$ (dashed red) and spontaneous work extracted $W$ (dash-dotted blue) as a function of $\theta$ for initial state preparation $\rho(0) = (1-p) \dyad{+_\theta} + p\dyad{-_\theta}$. Upper panel: $p=1/2$, middle panel: $p=1/4$, lower panel: $p=0$. Insets: Husimi function of the emitted field ${\cal Q}_\theta(\alpha)= \langle \alpha \dyad{\psi_\text{out}(\theta)} \alpha \rangle)$ for $\theta = 0, \pi/2, \pi$ and $p=0$. } 
\end{figure}

To further explore the role of quantum coherence, we focus on the scenario (ii). The battery is initially empty $\bar{N} = 0$, which corresponds to the spontaneous regime of light-matter interaction. From now on the considered scenarios are fully autonomous, i.e. do not require an external control of the qubit-battery coupling time. Integrating Eq.~\eqref{eqW} with $\dot{N}=0$ yields $W = \hbar \omega_0 s^2(0)$, revealing a fundamental and so far overlooked relation between work and coherence. Therefore as soon as $s(0) \neq 0$, a non-negligible fraction of the qubit's ergotropy ${\cal W}(0)$ can be spontaneously released as work, even though it is a dissipative mechanism. The work extracted $W$ is plotted in Fig.~\ref{fig3} together with the qubit's initial energy ${\cal E}(0)$ and ergotropy ${\cal W}(0)$ as a function of $\theta$ for $p=1/2,1/4,0$. We verify that $W\leq {\cal W}(0) \leq {\cal E}(0)$. $W$ is maximized for $\theta=\pi/2$ and $p=0$ which corresponds to the maximal initial coherence $s(0)=1/2$. On the opposite, $W$ vanishes for $\theta=\pi$, which corresponds to the case of a single photon source. Conversely as it appears on the figure, $\eta$ tends to $1$ in the limit $\theta \rightarrow 0$. Therefore in the case of an initially empty battery, work and yield cannot be optimized simultaneously. \\

To get an intuitive interpretation of these behaviors, it is fruitful to consider the quantum state of light spontaneously emitted in the WG during the process (See insets of Fig.~\ref{fig3} for a graphical representation). For $p=0$ it reads $\ket{\psi_\text{out}(\theta)} = \cos(\theta/2) \ket{0} + \sin(\theta/2) \ket{1}$, where $\ket{n}$ are the $n$-photon Fock states in the mode defined as $\hat{b} = \sqrt{\gamma} \int_0^\tau \dd t\,  \hat{b}_\text{out}(t)$. By definition, work corresponds to the energy carried by the coherent component of $\ket{\psi_\text{out}(\theta)}$ of amplitude $\beta_\theta = \bra{\psi_\text{out}(\theta)} \hat{b} \ket{\psi_\text{out}(\theta)}$, such that $W/\hbar \omega_0 = \cos^2(\theta/2) \sin^2(\theta/2)$. This translates the fact that single photons have no phase, such that single photon sources do not produce any work. Conversely, the yield compares the work to the total energy carried by the field, i.e. $\hbar \omega_0 \bra{\psi_\text{out}(\theta)} \hat{b}^\dagger \hat{b} \ket{\psi_\text{out}(\theta)}$. This brings out $\eta = \cos^2(\theta/2)$. Thus $\eta$ measures the overlap between the emitted field and the vacuum state, and is all the larger as the extracted work is lower. \\

\begin{figure}[htb]
    \includegraphics[width=\linewidth]{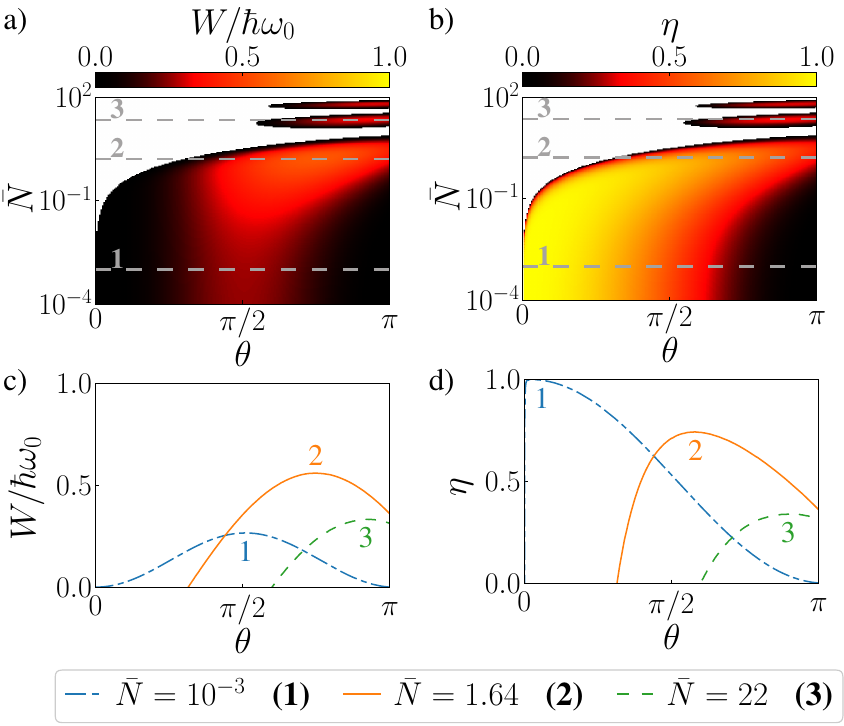}
    \caption{\label{fig4}
      Pulsed regime of work extraction (Case (iii), see text). \textbf{(a)} Work extracted $W$ and \textbf{(b)} yield of the protocol $\eta$, for an initial qubit's state $\rho(0) = \dyad{+_\theta}$, as a function of the battery's charge $\bar{N}$ and $\theta$. White color indicates negative work extraction. \textbf{(c)} $W$ and \textbf{(d)} $\eta$ as a function of $\theta$ for three different charges $\bar{N}$ (See legend and text).}
\end{figure}

To fully characterize the interplay between the battery's charge and the initial qubit's coherence, we finally consider the scenario (iii) where work extraction is stimulated by a resonant wave packet of finite charge and duration. Namely, we take as the qubit's initial state $\rho(0) = \dyad{+_\theta}$, while the WG is filled with a square wave packet of duration $\tau =  \gamma^{-1}$. The work extracted $W$ and the process yield $\eta$ are plotted in Fig.~\ref{fig4}a and b as a function of the battery's charge $\bar{N}$ and the angle $\theta$. Three regimes can be observed. Large initial battery's charges ($\bar{N}\geq 10$) induce stimulated emission. The phase of the coherent field partially radiated by the qubit is set by the drive. No initial coherence is required, work extraction and yield are simultaneously optimized for $\theta = \pi$. The optimal conditions of the scenario (i) can be recovered, by taking $\bar{N} \rightarrow N_\text{int} \gg 1$ and $\tau \rightarrow \tau_\text{opt} \ll \gamma^{-1}$. In the opposite regime ($\bar{N}\ll 1$), the phase of the emitted field can only be set by the quantum phase of the initial qubit's state, which requires the injection of coherence in the first place. Yield (resp. work) is optimized for $\theta \rightarrow 0$ (resp. $\theta = \pi/2$). An optimal amount of extracted work $W/\hbar \omega_0=0.57$ is reached in the intermediate regime $\bar{N} = 1.64$ where both the qubit's coherence and the battery's charge contribute (See Fig.~\ref{fig4}c). By improving the mode matching between the input field and the qubit, pulse shaping allows increasing this amount up to $W/\hbar \omega_0 = 0.7$ (See Suppl.~\cite{Suppl}).\\
 
We have shown that quantum coherence and energy are complementary resources for work extraction in a quantum battery. Our scenario is dual to former studies where the working substance is an electromagnetic mode and the battery is made of ensembles of qubits \cite{Campisi2018, andolina_charger-mediated_2018, Andolina2019}, the limitation of the work extraction coming from residual correlations between the mode and the emitters. In the present case, the limitations come from the quantum nature of the working substance and the dissipation of heat by its quantum fluctuations. 

Interestingly, the experiments we propose can be realized in state of the art qubit-light interfaces. Such interfaces are key components of quantum networks and quantum communication technologies \cite{kimble_quantum_2008}. They are currently implemented in various platforms and involve either a direct coupling between the qubit and the WG, or a coupling mediated by a cavity. The two situations are modeled by our formalism.

Our findings go beyond the thermodynamical framework they are phrased in. Classical resources are needed to generate the unitary operations that control quantum systems \cite{Abah_2019}, e.g. to perform quantum gates \cite{miko17, Renato19}, interferometric measurements \cite{Bertet2001}, or unlock the ergotropy contained in a quantum state as presently studied. The framework we propose fundamentally allows measuring the energetic cost of classicality, and how quantum coherence can mitigate this cost.   \\

\begin{acknowledgments}
    It is a pleasure to thank G. M. Andolina, M.F. Santos, L. Lanco and P. Senellart for enlightening discussions. The authors acknowledge the J-P Aguilar Ph.D. grant from the CFM foundation, the Agence Nationale de la Recherche under the programme ``Investissements d'avenir" (ANR-15-IDEX-02), and under the Research Collaborative Project ``Qu-DICE" (ANR-PRC-CES47). 
\end{acknowledgments}

%

\onecolumngrid
\vspace{\columnsep}
\setcounter{section}{0}
\setcounter{equation}{0}
\setcounter{figure}{0}

\setlength{\abovedisplayskip}{8pt}
\setlength{\belowdisplayskip}{8pt}
\setlength{\abovedisplayshortskip}{5pt}
\setlength{\belowdisplayshortskip}{5pt}

\renewcommand{\thefigure}{S\arabic{figure}}
\renewcommand{\theequation}{S\arabic{equation}}
\clearpage
\begin{center}
    \textbf{\large Supplemental material: The energetic cost of work extraction}
     \title{Supplemental material: The energetic cost of work extraction}
    \author{Juliette Monsel}
    \affiliation{Univ. Grenoble Alpes, CNRS, Grenoble INP, Institut N\'eel, 38000 Grenoble, France}
    \author{Marco Fellous-Asiani}
    \affiliation{Univ. Grenoble Alpes, CNRS, Grenoble INP, Institut N\'eel, 38000 Grenoble, France}
    \author{Benjamin Huard}
    \affiliation{Univ Lyon, ENS de Lyon, Univ Claude Bernard, CNRS, Laboratoire de Physique, Lyon, France}
    \author{Alexia Auff\`eves}
    \email{alexia.auffeves@neel.cnrs.fr}
    \affiliation{Univ. Grenoble Alpes, CNRS, Grenoble INP, Institut N\'eel, 38000 Grenoble, France}
\end{center}

\section{I.\quad Input-output formalism}

\subsection{A.\quad Model}
In the main text, we consider a qubit of ground and excited states denoted $\ket{g}$ and $\ket{e}$ and of transition frequency $\omega_0$. This qubit is embedded in a wave guide, namely a one-dimensional reservoir of electromagnetic modes indexed by their frequency $\omega$ and characterized by the normalized density of modes $\rho(\omega)$. Denoting as $\hat{b}_\omega$ the corresponding lowering operators, the total Hamiltonian reads \cite{gardiner_quantum_2010, walls_quantum_2008}
\begin{equation}
\hat{H}_\text{tot} = \frac{\hbar\omega_0}{2} (\hat{\sigma}_z+\mathbf{1}) + \int_0^\infty \dd\omega \hbar\omega \rho(\omega) \hat{b}^\dagger_\omega \hat{b}_\omega + \ii\int_{0}^\infty \dd\omega    \rho(\omega) \frac{    \hbar g(\omega)}{2} (\hat{b}_\omega^\dagger \hat{\sigma}_- - \hat{\sigma}_+ \hat{b}_\omega)
\end{equation}
where $\hat{\sigma}_z = \dyad{e} - \dyad{g}$, $\hat{\sigma}_- = \ket{g}\bra{e}$ and $\hat{\sigma}_+ = \hat{\sigma}_-^\dagger$. Solving the evolution of the system in the Heisenberg picture and tracing over the field shows that the qubit's observables undergo a damping characterized by the rate $\gamma = \pi g^2(\omega_0)\rho(\omega_0)/2$. We have made the assumption that the modes of the waveguide constitute a \textit{bona fide} reservoir, namely that the width $\Delta \omega$ of the function $\rho(\omega)g(\omega)$ obeys $\Delta \omega \ll \gamma$. 
We define the input operator $\bin(t) =  \sqrt{2/\pi}\int_0^\infty\dd\omega \sqrt{\rho(\omega)} \hat{b}_\omega \e^{-\ii\omega t}$ which is related to the output operator $\bout(t)$ by the so-called input-output equation \cite{gardiner_quantum_2010, walls_quantum_2008, wiseman_quantum_2010, vool_introduction_2017}
\begin{equation}
\bout(t) = \bin(t) + \sqrt{\gamma}\hat{\sigma}_-(t). \label{bout-supp}
\end{equation} 
In the following, the input drive is chosen to be a coherent field, therefore we can replace $\bin(t)$ by the field's complex amplitude $\bbin(t)$ and obtain Eq.~\eqref{bout} from the main text.

\subsection{B.\quad Input and output powers}
The mean value of the input and output operators is expressed in units of the square root of a photon rate. Defining the dimensionless mode $\hat{B}=\bin /\sqrt{\gamma}$ allows introducing the input field state $\ket{\beta_\text{in}} = {\cal D}_{\hat{B}}(\beta_\text{in})\ket{0}$ where ${\cal D}_{\hat{a}}(\alpha) = \e^{\alpha^*\hat{a} - \alpha \hat{a}^\dagger}$ is the displacement operator in the mode $\hat{a}$ by the amount $\alpha$ and $\ket{0}$ is the vacuum state. 
The rate of photons impinging on the qubit is $\dot{N}(t) = \langle \bin^\dagger(t) \bin(t) \rangle$. Conversely, the operator accounting for the rate of propagating photons in the output field is $ \bout^\dagger(t) \bout(t)$ and verifies
\begin{equation}
\bout^\dagger(t) \bout(t) = \bin^\dagger(t) \bin(t) +\gamma \hat{\sigma}_+(t) \hat{\sigma}_-(t)+\sqrt{\gamma}\left(\bin^\dagger(t)\hat{\sigma}_-(t) + \hat{\sigma}_+(t)\bin(t)\right)
\end{equation}
yielding $\ev{\bout^\dagger(t)\bout(t)} = |\bbin(t)|^2 + \gamma P_e(t) +2\sqrt{\gamma}\Re(\bbin(t) s(t))$. In the following, we choose the phase of the input drive so that $\bbin(t) = \sqrt{\dot{N}(t)}\e^{-\ii\omega_0 t}$. $P_e$ denotes the population of the qubit's excited state and the input and output powers read $P_\text{in/out}(t) = \hbar\omega_0\ev{\hat{b}_\text{in/out}^\dagger(t)\hat{b}_\text{in/out}(t)}$. The Rabi frequency is defined as $\Omega(t) = 2\sqrt{\gamma}|\bbin(t)|$ and we recover the usual Hamiltonian for a driven qubit: $\hat{H}(t) = \frac{\hbar \omega_0}{2}(\hat{\sigma}_z+\mathbf{1})+ \ii\hbar \Omega(t)/2 (\hat{\sigma}_- \e^{\ii\omega_0 t} - \hat{\sigma}_+ \e^{-\ii\omega_0 t})$. Finally, the mean energy of the qubit, $\mathcal{E}(t) = \Tr[\rho(t) H(t)]$, reads
\begin{equation}
\mathcal{E}(t) = \hbar\omega_0 P_e(t) - \hbar\Omega\Im(s(t)\e^{\ii\omega_0 t}),
\end{equation}
and the evolution of the population and coherence is given by the Bloch equations
\begin{subequations}\label{Bloch}
    \begin{align}
    \dot{P}_e (t) &= -\gamma P_e(t) -\Omega(t)\Re(s(t)\e^{\ii \omega_0 t}),\\
    \dot{s}(t) &= - \left(\ii\omega_0 + \frac{\gamma}{2}\right)s(t) + \Omega(t)\e^{-\ii \omega_0 t}\left( P_e(t) -\frac{1}{2} \right).\label{eq_s}
    \end{align}
\end{subequations}
The qubit's initial state is
\begin{equation}
\rho(0)= p \dyad{-_\theta} + (1-p) \dyad{+_\theta}, \label{rho(0)}
\end{equation}
where $p \in [0,1/2]$, $\theta \in [0,\pi]$ with $\ket{+_\theta} = \sin(\theta/2) \ket{e} +\cos(\theta/2) \ket{g}$ and $\ket{-_\theta} = -\cos(\theta/2) \ket{e} + \sin(\theta/2) \ket{g}$.
Thus, $s(0)$ is real and, from Eq.~\eqref{eq_s}, we obtain $\Im(s(t)\e^{\ii\omega_0 t}) = \Im(s(0))\e^{-\gamma t / 2}$, so we have $\Im(s(t)\e^{\ii\omega_0 t})=0$ at any time. Therefore, $\mathcal{E}(t) = \hbar\omega_0 P_e(t)$ and, finally, we get $\dot{\mathcal{E}}(t) = \hbar\omega_0 \dot{P}_e(t)$ and $P_\text{out}(t) = P_\text{in}(t) - \dot{\mathcal{E}}(t)$ (Eq.~\eqref{Pout} from the main text). 

\subsection{C.\quad Evolution for a square pulse}\label{evol_square}
We consider a square pulse of duration $\tau$, i.e. the Rabi frequency $\Omega = 2\sqrt{\gamma \dot{N}}$ is constant and non-zero on the time interval $[0, \tau]$.
Calling the coherence in interacting picture $\bar{s}(t)=\e^{\ii \omega_0 t} s(t)$, the Bloch equations \eqref{Bloch} become, for $t \in [0, \tau]$,
\begin{subequations}\label{Bloch2}
    \begin{align}
    \dot{P}_e(t) &= -\gamma P_e(t)  - \Omega  \bar{s}(t), \\
    \dot{\bar{s}}(t)  &= -\frac{\gamma}{2}\bar{s}(t)  +\Omega \left(P_e(t)  - \frac{1}{2}\right).
    \end{align}
\end{subequations}
From them, we derive the following equation for $\bar{s}(t)$:
\begin{equation}
\ddot{\bar{s}} + \frac{3\gamma}{2}\dot{\bar{s}} + \left(\Omega^2 + \frac{\gamma^2}{2}\right)\bar{s} = -\frac{\Omega\gamma}{2}.
\end{equation}
It has different solutions depending on how the Rabi frequency compares to the spontaneous emission rate. We denote $\epsilon=\frac{\gamma}{\Omega}$. Excluding the limit case $\epsilon=4$, we find:
\begin{itemize}
    \item If $\epsilon > 4$: exponentially decaying solution
    \begin{equation}
    \bar{s}(t) = \e^{-\frac{3\gamma}{4}t}\left(A \cosh(Dt) + B\sinh(Dt)\right) +C.
    \label{coherences_positive_delta}
    \end{equation}
    \item If $\epsilon < 4$: quasi-periodic solution
    \begin{equation}
    \bar{s}(t) = \e^{-\frac{3\gamma}{4}t}\left(A \cos(Dt) + B\sin(Dt)\right) +C.
    \label{coherences_negative_delta}
    \end{equation}
\end{itemize}
We have used the following notations:
\begin{align}
D &= \frac{\sqrt{|\gamma^2 - 16\Omega^2|}}{4},\\
B &= \frac{1}{D}\left[\left(\frac{1}{2} - p\right)\left(\frac{\gamma}{4}\sin\theta - \Omega\cos\theta\right) + \frac{3\gamma^2\Omega}{4(2\Omega^2 + \gamma^2)}\right],\\
C &= -\frac{\gamma \Omega}{2 \Omega^2+\gamma^2},\\
A &= \left(\frac{1}{2} - p\right)\sin\theta - C.
\end{align}
\subsection{D.\quad Work and heat rates}

As stated in the main text, work is identified with the energy carried by the coherent component of the field emitted by the qubit. Therefore, the work rate is given by
\begin{equation}
\dot{W}(t) = \hbar\omega_0 \big\vert\big\langle\bout(t)\big\rangle\big\vert^2  -\hbar\omega_0 \big\vert\big\langle\bin(t)\big\rangle\big\vert^2.
\end{equation}
Using the input-output relation \eqref{bout-supp}, we obtain $\dot{W}(t)  = \hbar \omega_0(\gamma |s(t)|^2 + \Omega \Re(s(t)\e^{\ii \omega_0 t}))$ (Eq.~\eqref{eqW} from the main text).
The heat is associated with the incoherent component of the emitted field. Using energy conservation, we obtain the heat rate
\begin{align}
\dot{Q}(t) &= -\dot{\mathcal{E}}(t) - \dot{W}(t) \nonumber\\
&= \hbar \omega_0 \gamma (P_e(t) - |s(t)|^2),
\end{align} 
which corresponds to Eq.~\eqref{eqQ} from the main text.

\section{II.\quad Tight bound for work extraction}   
In any thermodynamical transformation, the amount of extracted work is naturally bounded by the energy change of the physical system under study. It is the purpose of this Section to show that the ergotropy of the system's initial state provides a tighter bound, namely,  ${\cal E}(0) \geq {\cal W}(0) \geq W$.\\

By definition, the ergotropy is the maximum amount of work that can be extracted from a quantum state by unitary operations. Conversely, work is usually defined as the system's energy change under unitary evolution. Therefore the above mentioned inequality is naturally satisfied. In the present situation, we have extended the concept of work that is now identified with the energy carried by the coherent component of the battery's state. This new scenario calls for a dedicated demonstration.
\subsection{A.\quad Expression of the extracted work and ergotropy}

The energy of the initial state $\rho(0)$ (Eq.~\eqref{rho(0)}) reads
\begin{equation}
\E(0) = p \E_{\ket{-_{\theta}}} + (1- p)\E_{\ket{+_{\theta}}},
\end{equation}
where we have defined $\E_{\ket{\pm_{\theta}}} = \frac{\hbar\omega_0}{2} (1 \mp \cos\theta)$ the energies of the states $\ket{\pm_{\theta}}$.
These energies are linked by
\begin{equation}
\left( \E_{\ket{-_{\theta}}}- \frac{\hbar \omega_0}{2} \right)=- \left( \E_{\ket{+_{\theta}}}-\frac{\hbar \omega_0}{2} \right)
\end{equation}
since the states are diametrically opposed on the Bloch sphere. This leads to
\begin{equation}
\E(0)=(1-2p)\E_{\ket{+_{\theta}}} + p\hbar \omega_0.
\end{equation}
We call $R$ the unitary (rotation) we do to extract work. Ergotropy is the maximum amount of extractable work, so we are looking for the $R$ that maximizes the extracted work. After the interaction, the final state has the expression
\begin{equation}
\rho(\tau)=(1-p) R \ket{+_{\theta}}\bra{+_{\theta}}R^{\dagger} + p R \ket{-_{\theta}}\bra{-_{\theta}} R^{\dagger}.
\end{equation}
The energy of the final state thus has the expression
$\E(\tau)=(1-2p)\E_{R\ket{+_{\theta}}}+p \hbar \omega_0$,
and the extracted work is
\begin{align}
W &=\E(0)-\E(\tau) \nonumber\\
&=(1-2p)\left(\E_{\ket{+_{\theta}}}-\E_{R\ket{+_{\theta}}}\right).
\end{align}
Thus, maximizing the extracted work is equivalent to consider the rotation minimizing the energy of $R\ket{+_{\theta}}$, thus putting it in the ground state. It finally gives us
\begin{align}
\mathcal{W}(0) &=(1-2p)\E_{\ket{+_{\theta}}} \nonumber\\&=\hbar \omega_0 (1-2p) \sin^2(\theta/2).
\end{align}
This expression for $\mathcal{W}(0)$ can also be obtained by applying the general formula of ergotropy given in \cite{Allahverdyan_2004-supp}.

\subsection{B.\quad Bounding work with ergotropy}
We can split the work into its unitary and non unitary part via $W=W_\text{stim}+W_\text{sp}$.
\begin{align}
W_\text{stim} &= \hbar\omega_0 \Omega \int_{0}^{\tau} \dd t  \bar{s}(t),\label{Wstim}\\
W_\text{sp} &= \hbar\omega_0\gamma \int_{0}^{\tau}\dd t  \bar{s}^2(t) + \hbar\omega_0\bar{s}^2(\tau)\label{Wsp}.
\end{align}
$W_\text{stim}$ is associated with the resonant drive and therefore vanishes when the pulse ends, while $W_\text{sp}$ corresponds to the contribution of the spontaneous emission to the extracted work.\\

When the process is purely unitary ($\gamma=0$), $W=W_\text{stim}$ which is lower than the ergotropy by definition of this quantity. When the process is non unitary ($\gamma \neq 0$), $W=W_\text{stim}+W_\text{sp}$. In this regime, not only there is the additional term $W_\text{sp}$ that contributes, but the expression of the coherence also changes drastically, see Eqs.~\eqref{coherences_negative_delta} and \eqref{coherences_positive_delta}. Then, it is not easy at first sight to know if more work than ergotropy could be extracted.\\

We will now consider three specific regimes before going to the general one.\\
\paragraph{a.\quad Stimulated emission regime $\gamma \ll \Omega$:} We consider the case (iii) of a square pulse of duration $\tau$. During the pulse, since $\gamma \gg \Omega$, we can neglect spontaneous emission. But after the pulse, part of the spontaneous emission contributes to the work, therefore using Eqs.~\eqref{Wstim} and \eqref{Wsp}, we obtain
\begin{align}
W_\text{stim} &= \hbar\omega_0\left(\frac{1}{2} - p\right)(\cos(\theta - \Omega\tau) -\cos(\theta)),\\
W_\text{sp} &= \hbar\omega_0\bar{s}^2(\tau)\nonumber\\
&=\hbar\omega_0\left(\frac{1}{2} - p\right)^2\sin[2](\theta - \Omega\tau).
\end{align}
Therefore, 
\begin{equation}
W - \mathcal{W}(0) = \hbar\omega_0\left(\frac{1}{2} - p\right)\left[-p -\frac{1}{2} - \left(\frac{1}{2} - p\right)\cos[2](\theta - \Omega\tau) + \cos(\theta - \Omega\tau)\right],
\end{equation}
and by solving the equation $-p -\frac{1}{2} - \left(\frac{1}{2} - p\right)x^2 + x = 0$, we can see that $W - \mathcal{W}(0) \le 0$.\\
\paragraph{b.\quad Pure state:} In this case, $\rho(0) = \dyad{+_\theta}$ and therefore $\mathcal{W}(0) = \mathcal{E}(0)$.
The first law gives $-\Delta\mathcal{E} = W + Q$, with $\Delta\mathcal{E}$ the energy variation of the qubit. 
$-\Delta\mathcal{E} \le \mathcal{E}(0)$ and since the environment of the qubit is at zero temperature, the emitted heat $Q$ is non-negative, therefore $W \le \mathcal{W}(0)$.\\

\paragraph{c.\quad Spontaneous regime $\Omega = 0$:} This is the case (ii) from the main text and the evolution of the qubit (Eqs.~\eqref{Bloch2}) reads
\begin{align}
\dot{P}_e &= -\gamma P_e, \\
\dot{\bar{s}} &= -\frac{\gamma}{2}\bar{s},
\end{align}
Therefore, by solving the above equations, we obtain
\begin{align}
P_e(t) &= \left(\frac{1}{2} + \left(p - \frac{1}{2}\right)\cos\theta\right)\e^{-\gamma t},\\
\bar{s}(t) &=  \left(\frac{1}{2} - p\right)\sin\theta \e^{-\frac{\gamma}{2}t},
\end{align}
and the extracted work reads
$W = \hbar\omega_0 \left(\frac{1}{2} - p\right)^2\sin^2\theta$.
Then, we have
\begin{equation}
\mathcal{W}(0) - W = \hbar\omega_0\left(\frac{1}{2} - p\right)\left[(1 - \cos\theta) - \left(\frac{1}{2} - p\right)\sin^2\theta \right],
\end{equation}
so
\begin{equation}
\mathcal{W}(0) - W \ge  \hbar\omega_0\left(\frac{1}{2} - p\right)\left[1 - \cos\theta - \frac{1}{2}\sin^2\theta \right]\ge 0.
\end{equation}\\

\paragraph{d.\quad Square pulse:} We now consider the case (iii) of the main text for arbitrary values of $\Omega$ and $\gamma$.
In full generality, the work depends on several variables: $p,\theta,\gamma,\Omega,\tau$. The first goal is to simplify the problem by being able to remove some of the variables. This is the goal of the following two points.

\subparagraph{Property 1} \textit{The time leading to optimal work verifies $s(\tau_\text{opt})=0$.}\\

We will remove a first variable from the problem by determining $\tau_\text{opt}$ as a function of the other parameters. 
The work extraction will be optimum for $\partial_\tau W =0$, which leads to
\begin{equation}
\bar{s}(\tau) \left( \Omega +\gamma \bar{s}(\tau)+2 \partial_\tau\bar{s}(\tau)\right)=0.
\end{equation}
The first solution is $\bar{s}(\tau)=0$, the second one is $\Omega +\gamma \bar{s}(\tau)+2  \partial_\tau\bar{s}(\tau)=0$. Using the Bloch equations, this last condition becomes $\Omega P_e(\tau)=0$,
which correspond to the qubit in the ground state (thus included in the other solution). Therefore, the optimal work takes the form
\begin{equation}
W_\text{opt}=\hbar \omega_0 \int_0^{\tau_\text{opt}} \dd t \left( \Omega \bar{s}(t)+\gamma \bar{s}^2(t) \right).
\label{opt_work}
\end{equation}   

\subparagraph{Property 2} \textit{The optimal work is a function of $\gamma/\Omega$.}\\

We need to prove that that $A$, $B$, $C$ and $D \tau_\text{opt}$ (from Section \ref{evol_square}) only depend on the ratio $\gamma/\Omega$. Indeed, if this is true, as integrating Eq.~\eqref{opt_work} will only involve functions evaluated in $D \tau_\text{opt}$, and the terms $A$, $B$, $C$: $W_\text{opt}$ will depend on the ratio $\gamma/\Omega$ and not on those variables separately. Doing so, a second variable will be removed from the problem.\\

We had defined $\epsilon=\gamma/\Omega$, so we rewrite $\gamma=\epsilon \Omega$. It gives us:
\begin{equation}
C(\gamma=\epsilon \Omega,\Omega)=-\frac{(\epsilon \Omega)\Omega}{2\Omega^2+(\epsilon \Omega)^2}=-\frac{\epsilon}{2+\epsilon^2}
\end{equation}
Thus, this coefficient indeed only depends on $\epsilon$. The same proof holds for the coefficient $A$. Now, $B$ involves terms like $\gamma/D$ , $\Omega/D$ or $\frac{3 \gamma^2 \Omega}{4D(2\Omega^2+\gamma^2)}$. We will only show that $\Omega/D$ depends on $\epsilon$ only as the proof is similar for the others:
\begin{equation}
\frac{\Omega}{D}=\frac{4 \Omega}{\Omega \sqrt{|\epsilon^2-16|}}=\frac{4}{\sqrt{|\epsilon^2-16|}}
\end{equation}
At this point, we thus have: $A(\epsilon)$, $B(\epsilon)$, $C(\epsilon)$. We now need to prove that $D \tau_\text{opt}$ is also a function of $\epsilon$ only.\\

$\tau_\text{opt}$ is defined through $\bar{s}(\tau_\text{opt})=0$. For $\epsilon >4$, it leads to
\begin{equation}
0=\e^{-\frac{3\gamma}{4}\tau_\text{opt}}\left(A(\epsilon) \cosh(D\tau_\text{opt}) + B(\epsilon)\sinh(D\tau_\text{opt})\right) +C(\epsilon).
\end{equation}
This equation can be formally rewritten in the form
\begin{equation}
-C(\epsilon)=F_1(\epsilon)\e^{\Omega \tau_\text{opt}\left(-3/4 \epsilon+D/\Omega \right)}+F_2(\epsilon)\e^{\Omega \tau_\text{opt} \left(-3/4 \epsilon-D/\Omega \right)}.
\end{equation}
Where $F_1$ and $F_2$ are function only depending $\epsilon$ that we don't explicitly write. We previously showed that $\frac{D}{\Omega}$ is indeed a function of $\epsilon$. Thus, we can define the function $f_{\pm}(\epsilon)=-3/4 \epsilon \pm D/\Omega$ and we have
\begin{equation}
-C(\epsilon)=F_1(\epsilon)\e^{\left(\Omega \tau_\text{opt}\right)f_+(\epsilon)}+F_2(\epsilon)\e^{\left(\Omega \tau_\text{opt}\right) f_-(\epsilon)}.
\label{eq_tau_opt}
\end{equation} 
We will now show that $\bar{s}(\tau_\text{opt})=0$ implies $D \tau_\text{opt}$ is a function of $\epsilon$. Taking the derivative with respect to $\Omega$ of Eq.~\eqref{eq_tau_opt}, we obtain
\begin{align}
0&=F_1(\epsilon)f_+(\epsilon)\partial_{\Omega}\left[\left(\Omega \tau_\text{opt}\right)\right] \e^{\left(\Omega \tau_\text{opt}\right)f_+(\epsilon)}+F_2(\epsilon)f_-(\epsilon) \partial_{\Omega}\left[\left(\Omega \tau_\text{opt}\right)  \right] \e^{\left(\Omega \tau_\text{opt}\right) f_-(\epsilon)}.
\end{align}
At this point, either $\partial_{\Omega}\left[\left(\Omega \tau_\text{opt}\right)\right]=0$, and we have proven $\Omega \tau_\text{opt}$ only depends on $\epsilon$, either it is not and we can simplify by this term, which gives
\begin{equation}
-\frac{F_1(\epsilon)f_+(\epsilon)}{F_2(\epsilon)f_-(\epsilon)}=\e^{(\Omega \tau_\text{opt})\left(f_-(\epsilon)-f_+(\epsilon)\right)}.
\end{equation}
As the left hand side does not depend on $\Omega$, the right hand side should not as well. Thus $\Omega \tau_\text{opt}$ is necessarily a function of $\epsilon$ only. As $D \tau_\text{opt}=\frac{D}{\Omega}\Omega \tau_\text{opt}$ and we have proven that $\frac{D}{\Omega}$ depends only on $\epsilon$, we finally have that $D \tau_\text{opt}$ depends only on $\epsilon$. The proof is the same for $0<\epsilon<4$.

\begin{figure}[t!]
    \begin{minipage}[t]{0.49\textwidth}
        a)\hspace{0.5cm}\raisebox{-0.95\height}{\includegraphics[width=0.6\linewidth]{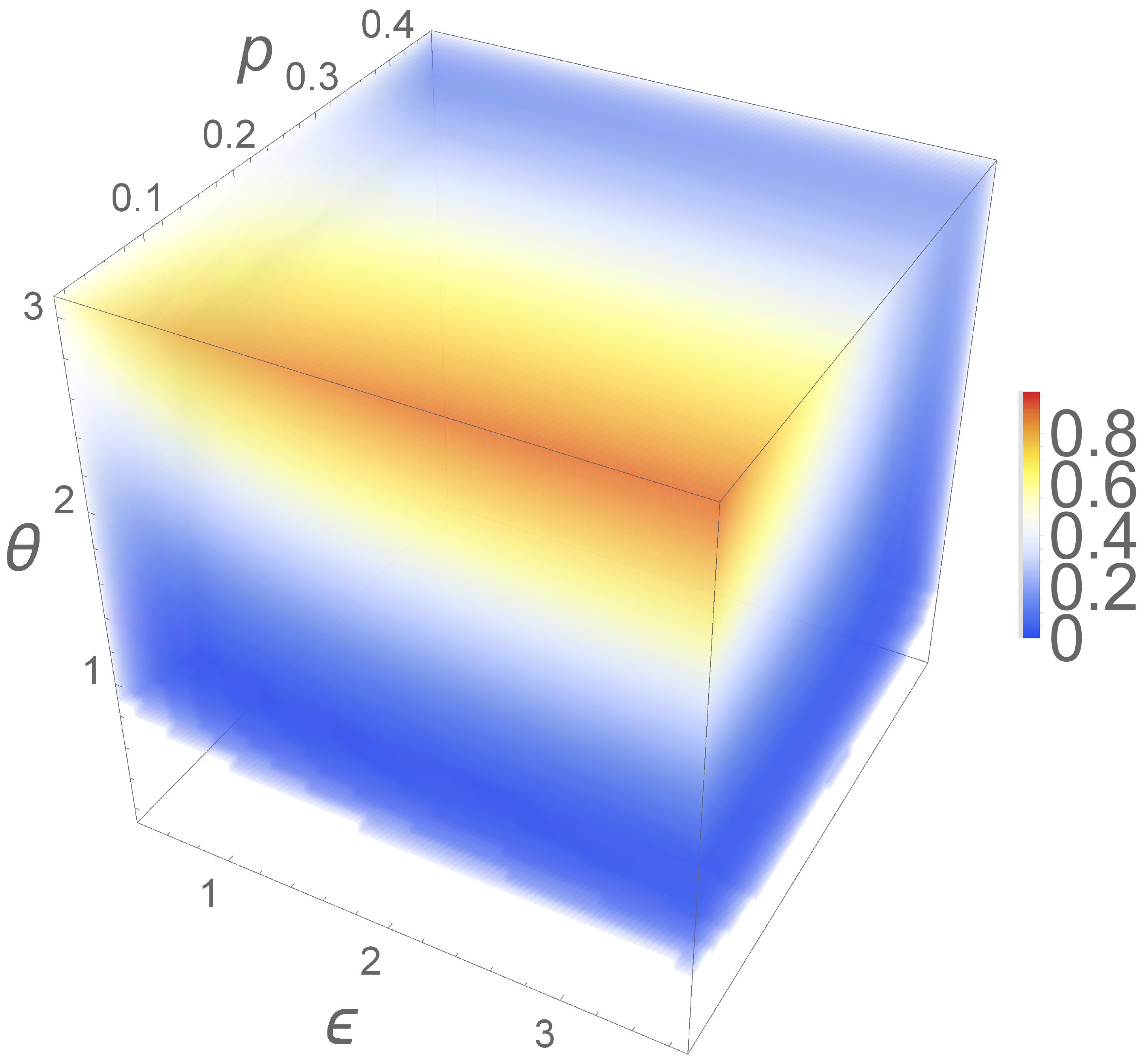}}
    \end{minipage}
    \begin{minipage}[t]{0.49\textwidth}
        b)\hspace{0.5cm}\raisebox{-0.95\height}{\includegraphics[width=0.6\linewidth]{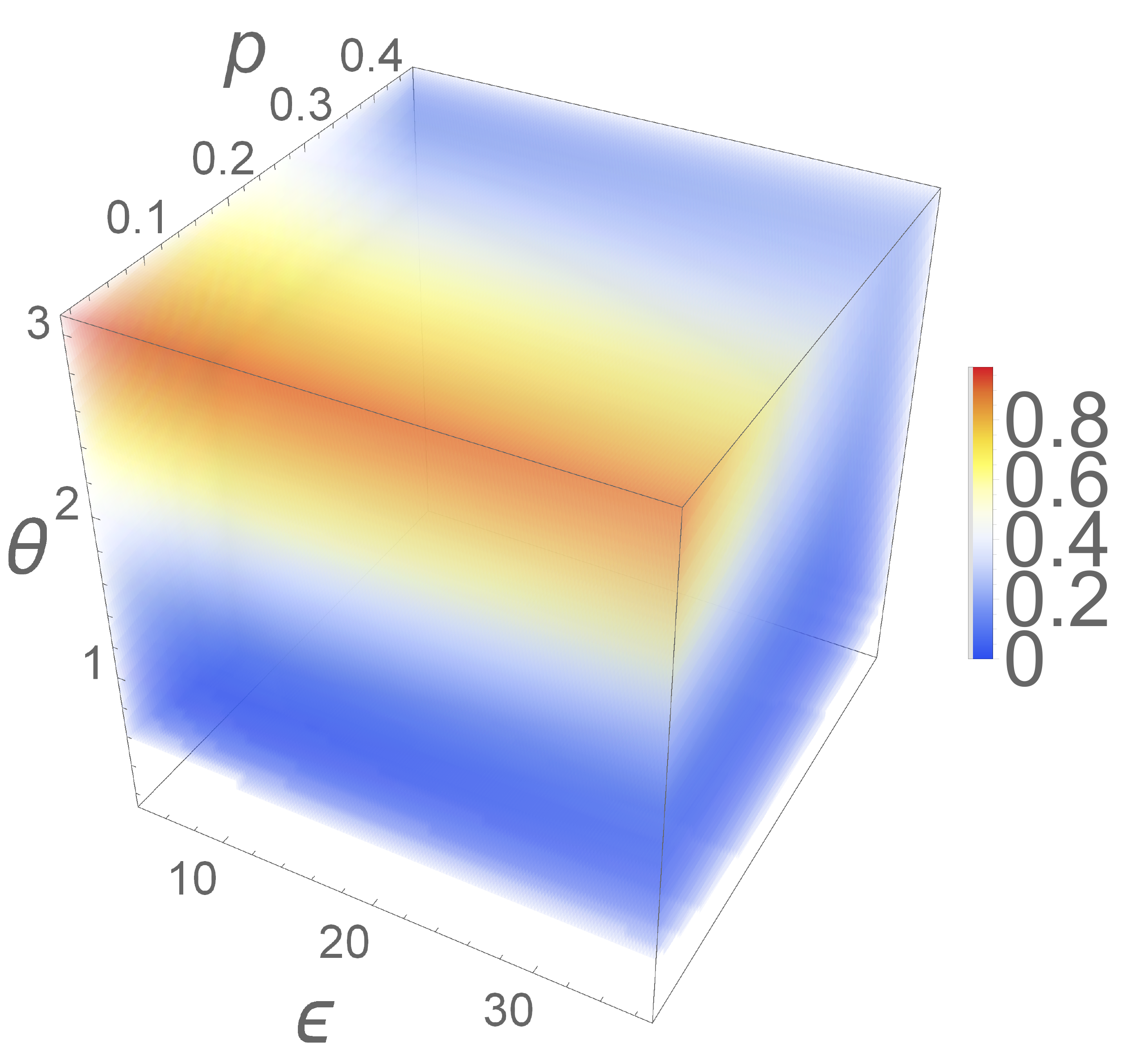}}
    \end{minipage}
    
    \caption{\label{fig_preuve_ergo}
        Difference between the ergotropy and the optimal extracted work, $\mathcal{W}(0)-W_\text{opt}(p,\theta,\epsilon)$,  \textbf{(a)}
        in the quasi-periodic regime ($0<\epsilon < 4$) and
        \textbf{(b)}
        in the exponentially decaying regime ($4<\epsilon$). The difference is always positive.
    }
\end{figure}  

\subparagraph{Final result} \textit{The optimal work is bounded by ergotropy.}\\

At this point, we have showed that
the optimal work extracted $W_\text{opt}$ is only a function of $p,\theta,\epsilon$, therefore
\begin{equation}
W(p,\theta,\Omega,\gamma,\tau) \leq W_\text{opt}(p,\theta,\epsilon).
\end{equation}
Finally, we have numerically checked that $W_\text{opt}(p,\theta,\epsilon) \leq \mathcal{W}(0)$. The quantity $\mathcal{W}(0)-W_\text{opt}$ is plotted in Fig.~\ref{fig_preuve_ergo} and is always positive. Since $W \leq W_\text{opt}$, we have $W \leq \mathcal{W}(0)$, ending the demonstration.

\section{III.\quad Optimization of the pulse shape}

\begin{figure}[hb!]
    \begin{minipage}[t]{0.49\linewidth}
        a)\raisebox{-\height}{\includegraphics[width=0.7\linewidth]{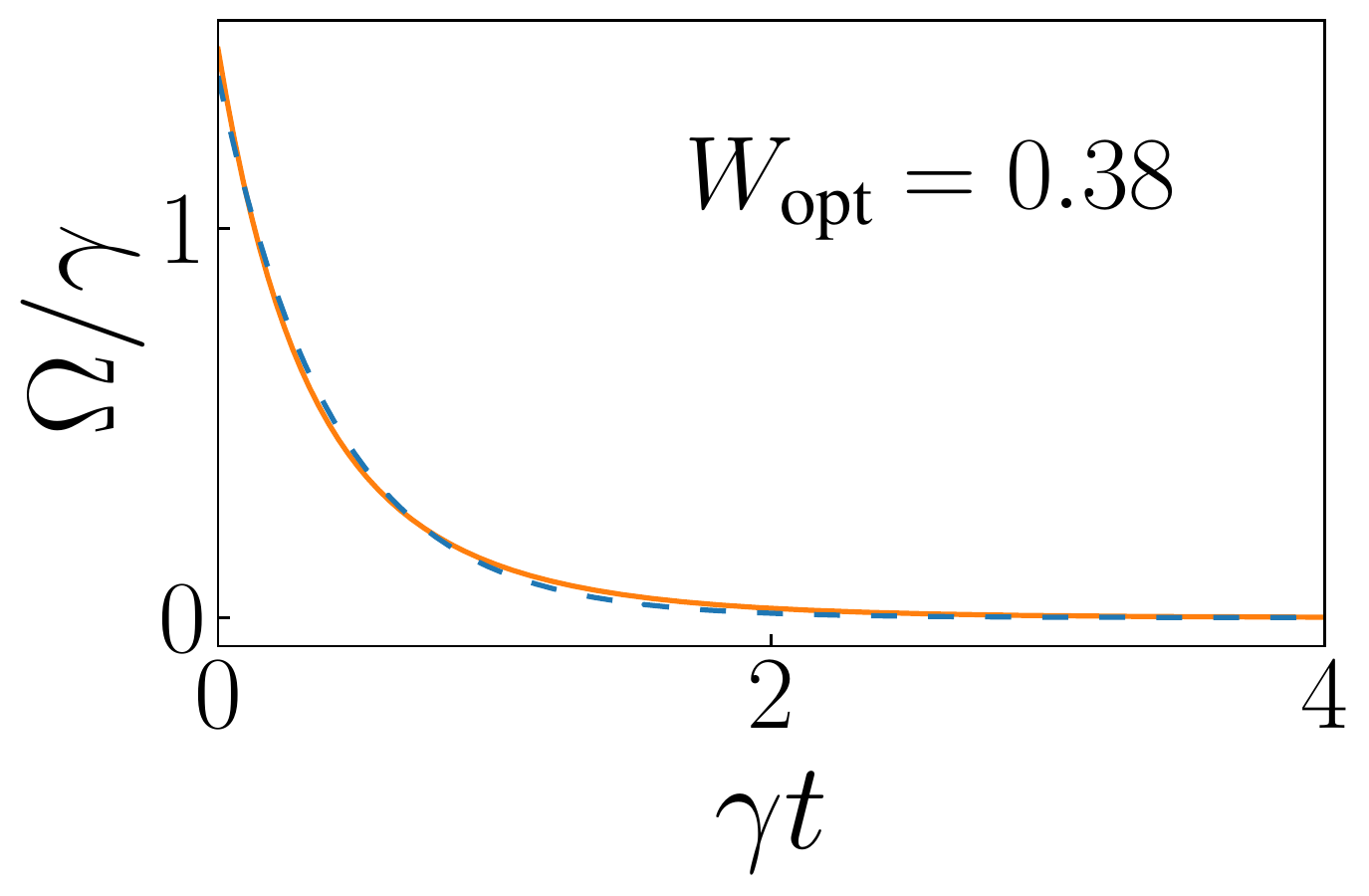}}
    \end{minipage}
    \begin{minipage}[t]{0.49\linewidth}
        b)\raisebox{-\height}{\includegraphics[width=0.7\linewidth]{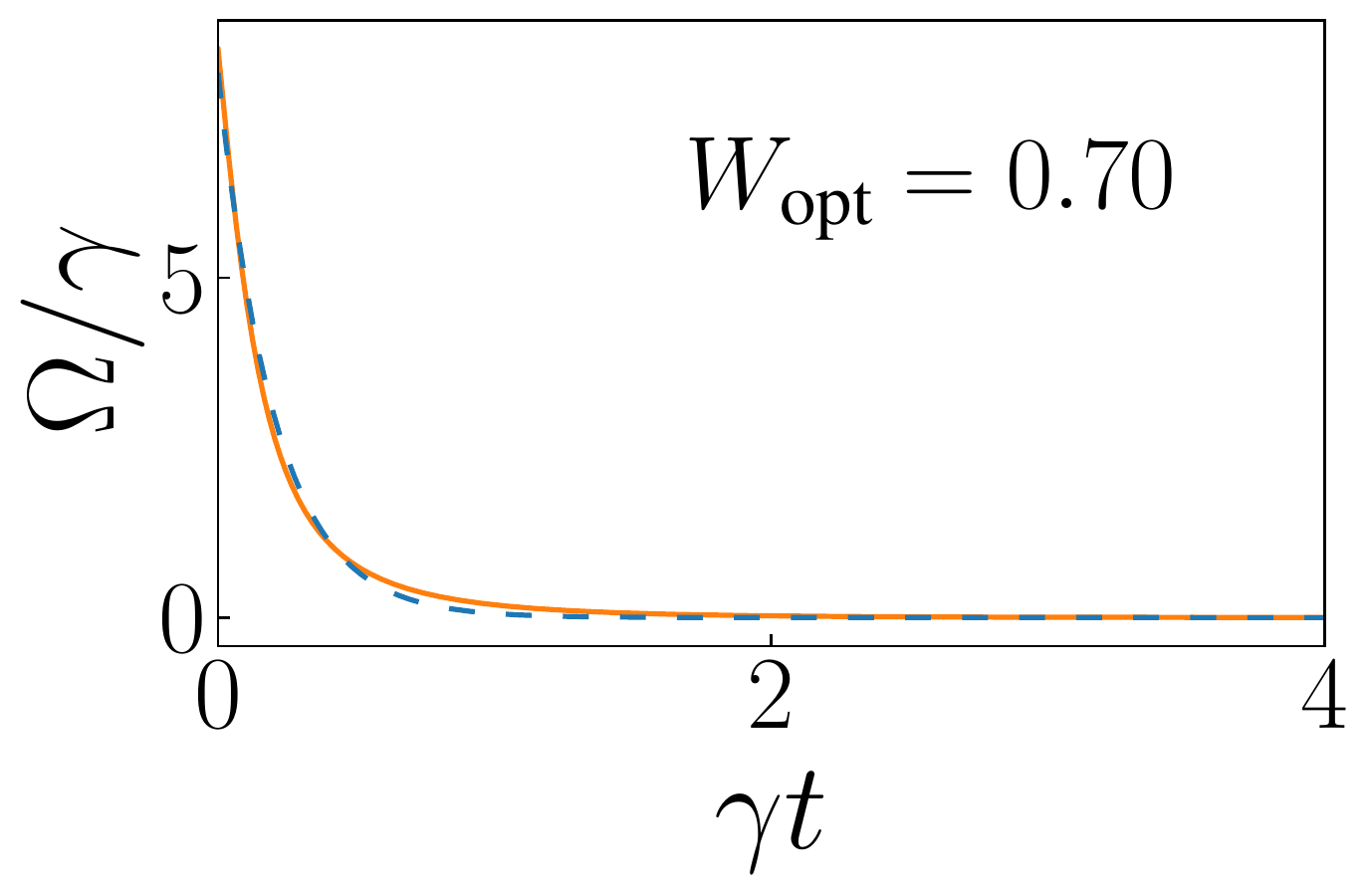}}
    \end{minipage}
    \includegraphics[scale=0.5]{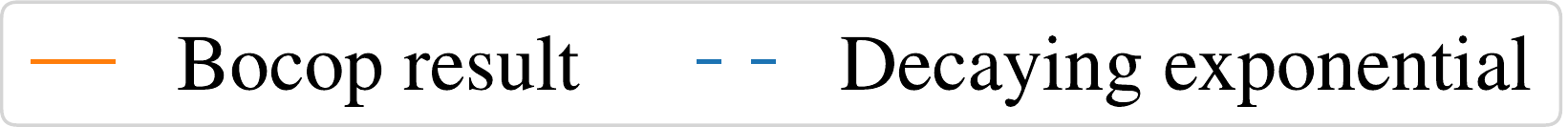}
    \caption{\label{fig-bocop}
        Comparison between the optimal pulse shape numerically computed by the software Bocop (solid orange) and the exponentially decaying pulse of optimal characteristic time $\tau_\text{opt}$ (dashed blue). Parameters: $p = 0$, $\gamma = 1$ GHz, in \textbf{(a)} $\bar{N} = 0.1$, $\theta = \pi/2$, $\gamma\tau_\text{opt} \simeq 0.41$ and in \textbf{(b)} $\bar{N} = 1.64$, $\theta = 3\pi/4$, $\gamma\tau_\text{opt} \simeq 0.20$.
    }
\end{figure}

We consider a qubit-field interaction of duration $T$ such that $T \gg \gamma^{-1}$. During this interaction, a fixed number of photons $\bar{N}$ is injected in the qubit with a pulse of Rabi frequency $\Omega(t)$. We want to find the pulse shape that maximizes the amount of work extracted for a given initial state $\rho(0)$ (Eq.~\eqref{rho(0)}). This is an optimal control problem where the quantity to maximize is $W = \int_0^T \dd t (\Omega(t)\Re(s(t)\e^{\ii\omega_0 t}) + \gamma \vert s(t)\vert^2)$ and the constraints are
\begin{align}
\int_0^T \dd t\, \Omega^2(t) &= 4\gamma \bar{N},\\
\dot{P}_e (t) &= -\gamma P_e(t) -\Omega(t)\Re(s(t)\e^{\ii \omega_0 t})\\
\dot{s}(t) &= - \left(\ii\omega_0 + \frac{\gamma}{2}\right)s(t) + \Omega(t)\e^{-\ii \omega_0 t}\left( P_e(t) -\frac{1}{2} \right),
\end{align}
We solved this problem numerically using Bocop \cite{Bocop} and compared its solution with an exponentially decaying pulse of Rabi frequency
\begin{equation}
\Omega_\text{exp}(t) = 2\sqrt{\frac{2\gamma \bar{N}}{\tau}}\e^{-t/\tau},
\end{equation} 
where $\tau$ is the characteristic time of the pulse.
The shape obtained for $\Omega(t)$ is plotted Fig.~\ref{fig-bocop} and can be superimposed on a decaying exponential of optimal characteristic time $\tau_\text{opt}$, namely the one maximizing work extraction. 
This effect was already observed in the
context of optimal irreversible stimulated emission \cite{Valente_2012} and
corresponds to the optimal mode matching between the drive
and the qubit.

\vspace{-1cm}    
 
%

\end{document}